\documentclass[utf8]{frontiersSCNS} 

\usepackage{url,hyperref,lineno,microtype,subcaption}
\usepackage[onehalfspacing]{setspace}
\usepackage{comment}
\usepackage{multirow}

\def\keyFont{\fontsize{8}{11}\helveticabold }
\def\firstAuthorLast{Jinqi Huang {et~al.}} 
\def\Authors{Jinqi Huang\,$^{1,*}$, Spyros Stathopoulos\,$^{1}$, Alex Serb\,$^{1}$, and Themis Prodromakis\,$^{1}$}
\def\Address{$^{1}$Centre for Electronics Frontiers, Electronics and Computer Science, University of Southampton}
\def\corrAuthor{Jinqi Huang}

\def\corrEmail{J.Huang@soton.ac.uk}

\begin{document}
\onecolumn
\firstpage{1}

\title[NeuroPack]{NeuroPack: An Algorithm-level Python-based Simulator for Memristor-empowered Neuro-inspired Computing} 

\author[\firstAuthorLast ]{\Authors} 
\address{} 
\correspondance{} 

\extraAuth{}

\maketitle

\begin{abstract}
\section{}
Emerging two terminal nanoscale memory devices, known as memristors, have over the past decade demonstrated great potential for implementing energy efficient neuro-inspired computing architectures. As a result, a wide-range of technologies have been developed that in turn are described via distinct empirical models. This diversity of technologies requires the establishment of versatile tools that can enable designers to translate memristors' attributes in novel neuro-inspired topologies. In this paper, we present NeuroPack, a modular, algorithm level Python-based simulation platform that can support studies of memristor neuro-inspired architectures for performing online learning or offline classification. The NeuroPack environment is designed with versatility being central, allowing the user to chose from a variety of neuron models, learning rules and memristors models. Its hierarchical structure, empowers NeuroPack to predict any memristor state changes and the corresponding neural network behavior across a variety of design decisions and user parameters options. The use of NeuroPack is demonstrated herein via an application example of performing handwritten digit classification with the MNIST dataset and an existing empirical model for metal-oxide memristors.
\tiny
 \keyFont{ \section{Keywords:} memristor, neuro-inspired computing, neuromorphic computing, neural networks, online learning, offline classification} 
\end{abstract}

\section{Introduction}

Over the last decade, neuro-inspired computing (NIC) has experienced an immense growth, manifesting itself in a range of advances across theory, hardware and infrastructure. Theoretical NIC has proposed a very wide range of artificial neural network (ANN) configurations, such as Convolutional neural networks \citep{cnn} and LSTMs \citep{lstm} that may operate at various levels of weight and signal quantisation \citep{signalQuant} and spanning across both spiking \citep{ANN} and non-spiking \citep{SNN} approaches. Evidently, this design process comprises multiple decision points that overall renders a very substantial and complex design space.

Simultaneously, research on novel hardware technologies has developed along multiple strands including fully digital architectures \citep{loihi, truenorth, spinnaker}, supra-threshold \citep{brainscales} and sub-threshold \citep{neurogrid} analogue architectures, with some more recent contenders \citep{pcm-ibm} utilising post-CMOS electronic components called "memristors" \citep{memristor}. This latter category is the focus of this work. Fundamentally, memristive devices are electrically tuneable (non-linear) resistors which have shown great promise at efficiently implementing the most numerous components found in neural networks: the synapses and mapping of their weights. Memristors feature the potential for extreme downscaling \citep{extremeScale}, back-end-of-line (BEOL) integrability \citep{memristor-cmos}, sub-ns switching capability \citep{highspeedrram} and very low switching energy \citep{ultralowcurrentrram}. Multiple families of memristive devices have been developed exploiting different electrochemical effects ranging from atomic-level effects in metal-oxides \citep{metal-oxide} and atomic switches \citep{AtomicSwitch}, to bulk crystallisation/amorphisation effects in phase-change memory devices \citep{pcm} and magneto-resistive effects in Spin-Torque-Transfer (STT) devices \citep{stt}, to name a few. Each of these families features its own idiosyncrasies in terms of electrical behaviour and correspondingly are described via distinct empirical models.

The broad interest in employing memristors within neuro-inspired hardware mainly stems from their Multi-bit storage \citep{multibit} capability and their simple architecture that can be tessellated in large arrays \citep{1t1r}. Those excellent features make memristors good candidates for multiply-accumulate operations \citep{hardwarebayesian} required in in-memory computing (IMC) applications. Moreover, the intrinsic properties of memristors are similar to those of biological synapses \citep{stdpandrram}. Inspired by this fact, designs such as \cite{unsupervised1} and \cite{unsupervised2} successfully applied memristors as synapses in online learning with spike-timing-dependent-plasticity \citep{stdp}, which is a learning rule inspired from biological NNs. Memristors have also been employed as components for NIC, from offline classification \citep{offline} to online learning \citep{online}. Along these lines, software-based simulation platform designed for memristor-based neuomorphic systems become significant for fast validation of design ideas and predicting device behaviour. 

Current simulators (e.g. MNSIM \citep{mnsim} and NeuroSim \citep{neurosim}) focus more on circuit-level designs, serving as tools either to simulate the behaviour of different hardware modules, or to estimate the performance of memristor-based neuromorphic hardware in integrated circuit designs. Sitting at a higher level of abstraction would be an "algorithm-level device-model-in-the-loop" simulator (or "algo-simulator" for short) designed to test functionally defined (as opposed to explicitly designed) circuits with memristive device models at algorithm level, e.g. performing specific online or offline learning tasks with memristors as synapses in spike-based NNs. Such tool would allow fast verification of design concepts before serious hardware design effort is committed, in essence answering the question: Is my design likely to function given knowledge on my memristive devices, assuming the rest of the circuit functions flawlessly? If yes, work can proceed to the next stage.

In this paper, we present NeuroPack: a simulator for memristor-based neuro-inspired computing at algorithm level. NeuroPack is a complete, hierarchical framework for simulating spiking-based neural networks, supporting various neuron models, learning rules, memristor models, memristor devices, neural networks, and different applications. Written in Python, it can be easily extended and customised by users, as will be shown. NeuroPack also integrates an empirical memristor model proposed by \citep{rrammodel}. Between processing algorithms and setting \& monitoring memristor states, there is the significant step of applying a pulse of specific voltage and duration to trigger a memristor state change corresponding to some desired weight change calculated from learning rules. NeuroPack integrates a module to convert desired weight changes to estimated stimulation pulse parameters for bridging this gap. In terms of applications, we use NeuroPack to demonstrate image classification on MNIST dataset in our 'Results' section. We also give result analysis for systems with different R tolerance, a parameter used in weight updating, and two biasing methods as examples to showcase that NeuroPack assists users to investigate how key design, device and architectural factors affect memristor-based neuromorphic computing systems. Finally, NeuroPack includes a built-in analysis tool with a user-friendly graphic user interface (GUI) for visualising and processing classification results. The main contributions of this work include:


\begin{enumerate}
    \item Developing an algo-simulator for memristor-powered neuro-inspired computing with selectable neuron and device models, as well as learning rules.
    \item Modelling memristor state changes in neuro-inspired computing tasks given user-defined memristor parameters.
    \item Converting expected weight changes prescribed by learning rules into parameters of bias pulses used for triggering memristor state changes in weight updating.
\end{enumerate}

The rest of paper is organised as follows: in section 2 we introduce the architecture of NeuroPack with core parts and the workflow. Section 3 demonstrates an example application of handwritten digit recognition in MNIST dataset performed in NeuroPack and section 4 summarises the paper.

\begin{figure}[t]
    \centering
    \includegraphics[width=1.0\linewidth]{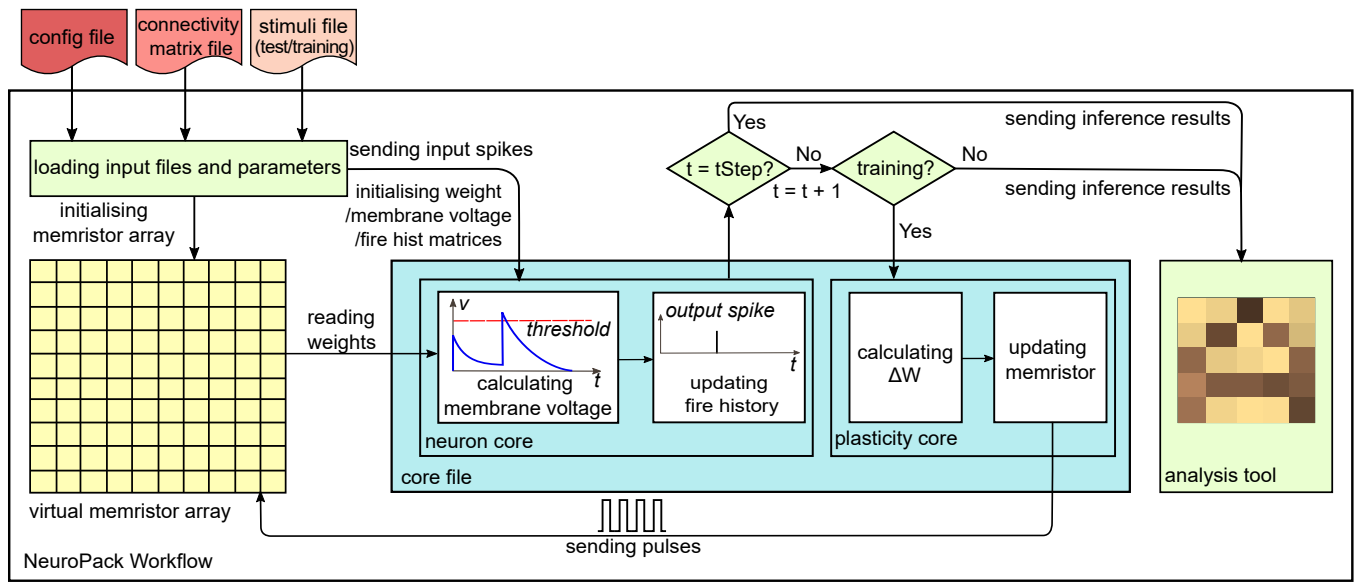}
    \caption{NeuroPack workflow. The system reads three configuration files: connectivity matrix, (neural) stimuli and config. The memristor model is embedded within the virtual memristor array, which initialises a number of memristor devices according to instructions in the configuration and connectivity matrix files. Neuron models and learning rules are placed in the core file. In "updating fire history" stage, there are three separate steps: calculating fire states assuming neurons fire 'freely'; adding network-level constraints; and updating the fire history matrix. "Calculating \(\Delta\)W" is supported by most learning rules except STDP. By replacing the core file, users can apply different neuron models and learning rules.}
    \label{fig:neuropack}
\end{figure}

\section{Methods}

\subsection{Design Overview}
NeuroPack is designed for predicting the outcomes of online learning or offline classification tasks under selectable neuron, plasticity and memristive device models, as well as for triggering and monitoring memristor state changes. To achieve those two tasks NeuroPack's workflow (see Figure \ref{fig:neuropack}) is divided into five parts: input file handling, virtual memristor array, neuron core, plasticity core, and analysis tool.

For input data handling, there are three input files that need to be generated from users: \emph{configuration file}, \emph{connectivity matrix file}, and \emph{stimuli file}. The \emph{configuration file} contains the main parameters for building up NNs (e.g. network size, NN depth, number of neurons for each layer etc), setting up neuron models (e.g. leakage, noise scale etc), and initialising memristor devices (e.g. up and bottom boundaries for memristor resistance). The \emph{connectivity matrix file} is used to define neuron connectivity and to map synapses to virtual memristors. The \emph{stimuli file} stores both input signals and output labels. When loaded in the NeuroPack main panel, input signals and output labels are split by checking if the neuron is an output neuron using the information provided by the \emph{configuration file}.

We now walk through the procedure for carrying out a classification task in a spiking network. First, the input signals to the neuron model are converted to spikes and saved in the \emph{stimuli file}. Training and test datasets are loaded separately. At each time step an input neuron can either be spiking (1) or silent (0). Next, the neuron core reads out the resistive states (RS) from a virtual memristor array, and calculates internal variables, such as membrane voltages, according to the selected neuron model. The new fire states are then calculated in two steps: firstly considering whether neurons are supposed to fire by checking if the membrane voltages surpass the threshold, and secondly adding network-level constraints (for example winner-take-all networks). When the current input stimuli belong to a training dataset, inference results are sent to the plasticity core to trigger weight update as per selected learning rule. When the test dataset is processed, plasticity updates are skipped.

Weight updates happen during the plasticity phase. For STDP, long-term potentiation (LTP) or long-term depression (LTD) "events" are directly applied to memristor devices based on a calculation taking nothing into account except for spike timing information. For other learning rules other types of information may be necessary. These can be made available to the system \emph{configuration file}. Importantly, there are two main conceptual ways of specifying plasticity events. The simpler way is to create a function that maps plasticity-relevant variables directly to pulsing parameters. Thereafter the physics of the memristor (correspondingly the response of the memristor model) will determine the actual resistive state change, which in turn will be reflected into a weight change via a resistance-to-weight mapping function. The more complex route involves mapping plasticity-relevant variable configurations to a requested weight change, and then searching the device model (a model is necessary for this approach) for a solution predicting that some set of pulse parameters will result in the required change. The same process will be repeated for the given number of epochs. Inference results as well as internal variable and parameters, including membrane voltages, fire history, and weights, can be sent to the built-in analysis tool for further visualisation and analysis. 

\subsection{Neuron Models and Learning Rules}

Neuron models and learning rules are placed in "neuron core" and "plasticity core" respectively in Figure \ref{fig:neuropack}. For some learning rules, such as backpropagation which requires calculation of error gradients, the equations depend on the neuron model. Therefore, neuron models must be treated as first-class objects. However, a simpler solution implemented in this tool is to ensure that each learning rule is paired with one neuron core and be placed together in the same "core" document. NeuroPack provides four example core files: leaky-integrate-and-fire neuron (LIF)\citep{lif} with STDP, leaky-integrate-and-fire neuron with backpropagation (BP)\citep{bp}, Tempotron \citep{tempotron}, Izhikevich neuron \citep{Izhikevich} and direct random target projection (DRTP) \citep{drtp}. Other neuron and plasticity rules can be customised according to users' needs by simply using existing example cores as standard templates and introducing user-defined cores. In this section, we provide one specific example, where we use a LIF neuron model and BP, to implement a fully-connected multi-layer spiking neural network with winner-take-all functionality \citep{wta}. 

\subsubsection{Leaky-integrate-and-fire neuron and Back-propagation}

Leaky-integrate-and-fire (LIF) is a simple and relatively computationally friendly neuron model. Most neuromorphic accelerators, with (e.g. \cite{memristor_snn_unsupervised}) or without memristors (e.g. \cite{digital_crossbar, algorithm_and_hardware}), use LIF neurons. LIF has been implemented in NeuroPack using the equations below adapted from the differential form \citep{originalLIF} by assuming discretised time steps: 

\begin{equation}
  \begin{gathered}  \label{eq:1}
        V_t = \sum Wx_t + \alpha V_{t-1}(1 - y_{t-1}) \\
        y_t = h(V_t - V_{th})   
    \end{gathered}
\end{equation}

Where \(V_t\) represents membrane voltage at time t, \(W\) is the weight, \(x_t\) is incoming spike to the neuron (considered as 1 or 0), $\alpha \in [0,1]$ is a leakage term, \(y_t\) is the output spike, \(h(x)\) is the step function and \(V_{th}\) is the neuron threshold. The equations describe that the neuron membrane voltage at any time step is determined by two parts: the weighted sum of incoming spikes at this time step and the membrane voltage at the last time step. If the membrane voltage surpasses the threshold, a spike is generated and the membrane voltage is reset. The cost function \(E\) at time step $t$ is then given by:

\begin{equation}    \label{eq:2}
    E(t) = \frac{1}{2N}\sum_{i = 0}^{N}(y_{i,t}-\hat{y}_{i,t})^2
\end{equation}    
where \(N\) is the number of output neurons, $i$ is the output neuron index, and \(\hat{y}_{i,t}\) is the correct firing state of output neuron \(i\) at time step $t$. Finally, weight changes are given by:

\begin{equation}    \label{eq:3}
    \begin{gathered}
        \Delta W_k = -\eta\delta_{k, t}x_{k, t}^T \\
        \delta_{k, t} = 
        \begin{cases}
        \frac{1}{N}(y_{k, t} - \hat{y}_t) \odot h'(V_{k, t} - V_{th}) & \text{if k = K}\\
        (W_{k+1, t}^T\delta_{k+1, t}) \odot h'(V_{k, t} - V_{th}) & \text{otherwise}
        \end{cases}
    \end{gathered}
\end{equation}
where \(\eta\) is the learning rate, \(k\) is the layer index, and \(K\) is the number of layers. \(W_k\) gives the weight matrix between layer \(k-1\) and \(k\). \(\odot\) means element-wise multiplication. \(\delta\) gives the error back-propagated from output neurons. The step function \(h(V_{t} - V_{th})\) is discontinuous, so the derivation is replaced by noise as suggested in \cite{backpropagation_with_noise}. For the full derivation, please refer to supplementary material. We note that the only potentially problematic variable is $\hat{y}_{i,t}$ which is provided in \emph{stimuli file}; everything else is accessible directly to NeuroPack. 

\subsubsection{Adding winner-take-all functionality}

We now introduce winner-take-all (WTA) functionality, which constrains the output layer to have at most one firing neuron at each time step. This is done by adding one softmax layer and making the neuron that has the largest softmax result fire:

\begin{equation}
    S_t = softmax(V_t \odot y_t) 
\end{equation}

The cost function correspondingly needs adjusted by taking a cross-entropy form:

\begin{equation}
    E = - \sum_{i = 0}^{N}\hat{y_{i,t}}ln(S_{i, t}) = -ln(S_{j,t})
\end{equation}

Where j is the index of the output neuron that should fire. Based on new cost function, weight changes are calculated as:

\begin{equation}
    \begin{gathered}
        \Delta W_k = -\eta\delta_{k, t}x_{k, t}^T \\
        \delta_{k, t} = 
        \begin{cases}
        (S_{t} - \hat{y_t}) \odot (y_t + V_t \odot h'(V_{k, t} - V_{th})) & \text{if k = K}\\
        (W_{k+1, t}^T\delta_{k+1, t}) \odot h'(V_{k, t} - V_{th}) & \text{otherwise}
        \end{cases}
    \end{gathered}
\end{equation}
as before application of WTA, all variables are accounted for; the freshly introduced softmax can be computed entirely within the same core file by simply adding network-level constraints. For the full derivation, please refer to supplementary material.

In summary, to configure a multi-layer spiking neural network with WTA using LIF neurons with BP in NeuroPack, parameters including learning rate, noise scale, threshold and leakage are loaded from the \emph{configuration file}. Input spikes encoded from stimuli files are fed to the "neuron core" in the core file. The "neuron core" reads the memristor states from the device array, calculates membrane voltages, and updates fire history during the inference phase. If training is enabled, inference results, ground truth information as well as internal variables are loaded into the "plasticity core" which then adjusts the memristive weights according to calculated weight changes, precisely as summarised in Figure \ref{fig:neuropack}.

\subsection{Memristor Model}
When using virtual memristive devices, as is the case in Figure \ref{fig:neuropack}, a memristive model has to be used for predicting memristor RS as a function of read-out voltage and RS changes in response to plasticity events. Here we use an empirical memristor model proposed by \cite{rrammodel}, but other user-defined models are also compatible with NeuroPack. With different values of parameters, this model has shown flexibility to model large ranges of memristor devices. The model expresses RS switching rate (\(\frac{dR}{dt}\)) as a function of initial RS and biasing voltage. The switching rate equation is reproduced here for convenience:

\begin{equation}\label{eq:switching}
    \frac{dR}{dt} = m(R, v) = 
     \begin{cases}
    A_p(-1+e^{\frac{|v|}{t_p}})(r_p(v) - R)^2& \text{if \( v >0,R<r_p(v)\)}  \\
    A_n(-1+e^{\frac{|v|}{t_n}})(R - r_n(v))^2& \text{if \(v \leq 0,R\geq r_n(v)\)}  
 \end{cases}
\end{equation}

where \(A_n\) and \(A_p\) are scaling factors. The $(e^{\frac{|v|}{t_{p,n}}}-1)$ term marks the main, exponential dependency of the switching rate on bias voltage with \(t_n\), \(t_p\) as fitting parameters extracted during modelling. Next, the last term encapsulates the dependence of the switching rate on the current resistive state with the aid of fitting parameters \(a_{0p}\), \(a_{1p}\), \(a_{0n}\), \(a_{1n}\). The main effect here is that the closer the RS is to the upper (lower) limit, the harder it is to continue pushing it further up (down), implementing "RS saturation". Finally, \(r_{p,n}(v)\) is a simple, 1st order polynomial helper function that helps capture the exact nature of the switching rate's dependency on current RS:

\begin{equation}
    r(v) = 
    \begin{cases}
    r_p(v) = a_{0p}+a_{1p}v & \text{if \(v > 0\)}\\
    r_n(v) = a_{0n}+a_{1n}v & \text{if \(v\leq 0\)}
    \end{cases}
\end{equation}

Whilst models may take different forms (e.g. charge-flux models \citep{charge-flux}), the fundamental condition that is observed is that the device model is always fed a series of time-wise discretised voltages and then must be able to calculate current and change in resistive state. Models that by default take current inputs are presently not supported. \(dt\) is treated as a pre-defined, fixed parameter and remains constant throughout simulation in the current implementation. NeuroPack organises memristor model instances in a virtual array with user-defined parameters as inputs by using a class \textbf{\text{ParametricDevice(*args)}} to create a device object and methods \textbf{\text{initialise(R)}} and \textbf{\text{step\_dt(V, dt)}} in the same class to set the current resistance and send a pulse with magnitude of \(V\) and pulsewidth of \(dt\) to the memristor device correspondingly. "Virtual memristor array" also provides methods such as \textbf{read(w,b)} and \textbf{pulse(w, b, v, pw)}, to read the RS of the device at a given wordline \textbf{w} and bitline \textbf{b} address within the virtual array and apply a pulse with magnitude of \textbf{v} and pulse-width \textbf{pw} to the device at said address correspondingly. It is these functions that neuron models and learning rules in the core files use to access memristor RS and trigger RS change. Using model parameter sets from different types of devices (different stacks or even devices with different underlying electrochemical mechanisms), as well as varying the parameters of the model in an exploratory fashion are great tools to gain an understanding of how different memristive devices can influence the outcome of basic machine learning algorithms. 

\subsection{Weight Mapping and Updating} 
Most memristor-based neuromorphic designs (e.g \cite{dotproduct} and \cite{insitu}) store weights as memristor conductance and apply incoming inputs as voltages across memristors, so that the multiplication results can be easily attained by measuring the current according to Ohm's law. Therefore, weights are mapped linearly to memristor conductance by default in NeuroPack, but users are allowed to define other mapping methods if needed. However, memristor being nonlinear devices, obtaining well-controlled weight changes is challenging because the resistive state after application of a triggering pulse depends both on the current resistance state and biasing parameters (typ. pulse voltage and duration). To deal with this issue, NeuroPack includes a module for calculating biasing parameters, expected to produce the desired weight changes. In this module, inputs are memristor model parameters for initialising a memristor device, current state, target state, \(dt\), and a list of tuples, each of which contains magnitude and pulsewidth to represent a pulse. 

Inside the module, a \textbf{ParametericDevice} object is created by passing memristor model parameters. Resulting resistance for each set of pulse parameters is calculated by calling \textbf{\text{step\_dt(V, dt)}} method. After all resulting resistance values are attained, distance to the resistance expected to be written to memristors is calculated, and the set of parameters that lead to shortest distance will be selected. The pseudo code of this module has be included in supplementary material. 

The weight update scheme using the pulse parameter selection module is as follows: to begin with, user-defined R-tolerance, which is defines as the maximum \(\frac{R_{new} - R_{expected}}{R_{expected}}\) that NeuroPack assumes the device state has converged, and maximum updating steps are loaded in Neuropack. After calculating the target resistance for a device, actual resistance is read and the resulting value is used as the initial resistance in the pulse parameter selection module. After attaining the set of parameters that will lead to closest resistance a pulse is sent and the new resistance is read. This predict-write-verify process is repeated until the calculated \(\frac{R_{new} - R_{expected}}{R_{expected}}\) is smaller than the R-tolerance or the maximum step number is reached.

\subsection{Customisation, usage scenarios and interface}

With Python as programming language, NeuroPack can be flexibly customised at both algorithm level and device level. At algorithm level, NeuroPack can apply user-defined neuron models and learning rules. As it is shown in Figure \ref{fig:neuropack}, neuron core and plasticity core are placed in the same \emph{core file}. In NeuroPack there are four example core files which will be explained in details in the next section. By writing and selecting user-defined core files, users can apply any customized neuron models and plasticity rules. At device level, users can easily set and change memristor parameters to describe different device characteristics.

\begin{figure}[t]
    \centering
    \includegraphics[width=1.0\linewidth]{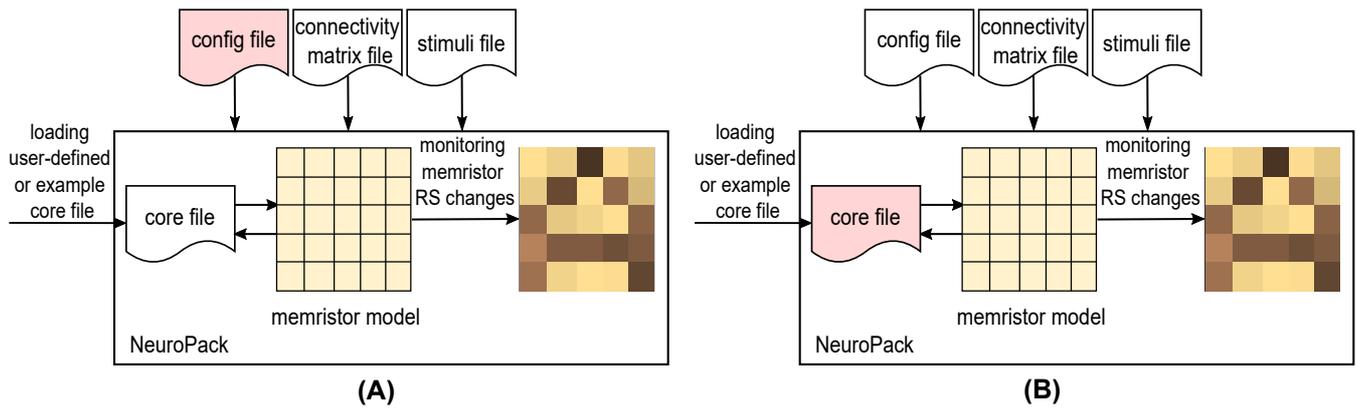}
    \caption[Usage scenarios for NeuroPack]{Usage scenarios for NeuroPack. Pink icons show targeted part of design for exploration in different scenarios.}
    \label{fig:scenario}
\end{figure}

There are two main usage scenarios for NeuroPack. One is as a supporting tool to explore how different parameter values and settings affect classification performance in NIC tasks (Figure \ref{fig:scenario} (A)). In this scenario, Users can load parameters in the configuration file with different values to monitor how memristor devices behave differently, or to investigate how classification accuracy or error rate is affected. Another usage scenario is to use NeuroPack to test and validate NN algorithms, including neuron models and learning rules (Figure \ref{fig:scenario} (B)). In this scenario, a user-defined core file or an example template is loaded to perform NN simulations. Users can then quickly test the algorithm and validate the idea by visualising and displaying memristor device state and other NN key variables (such as membrane voltage).    

The visualisation and analysis tool in NeuroPack provides a user-friendly GUI to display inference results. This tool is separated from the main panel of NeuroPack. When executing a classification task, NeuroPack saves variables in every epoch including membrane voltages, input stimuli, fire history, output errors, weights as measured from the memristors, etc in a Numpy (.npz) file. Array-related variables such as weights are displayed in a color map, and neuron-related variables, for example membrane voltages and fire history, are visualised in curves to show the changing tendencies.

\section{RESULTS}

\begin{figure}[t]
    \centering
    \includegraphics[width=0.98\linewidth]{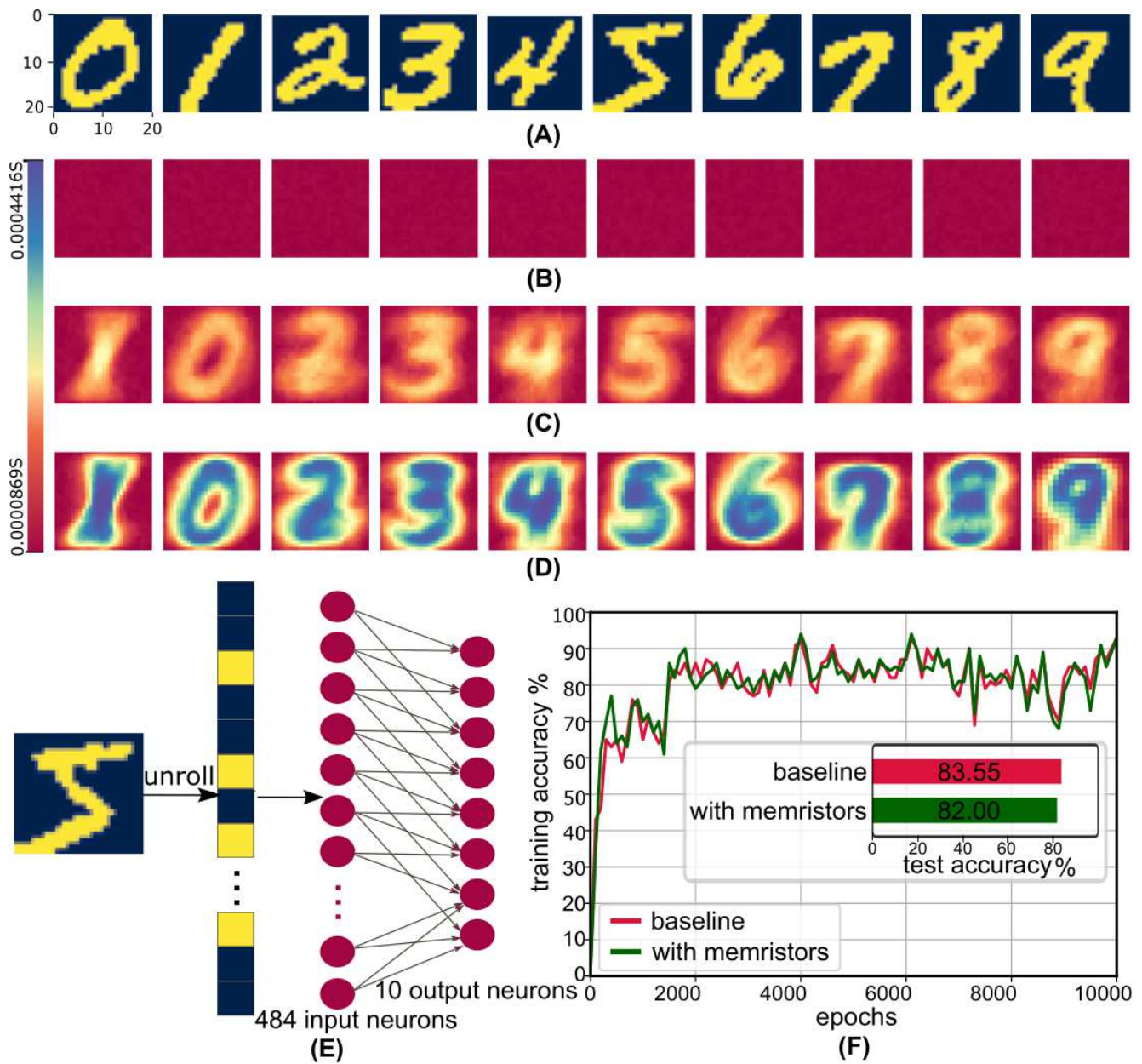}
    \caption{Image recognition task with TiOx memristor-based neural network simulation in NeuroPack. \textbf{(A)} Binary handwritten digits from MNIST dataset cropped to \(22 \times 22\). Dark blue and yellow represent weights 0 and 1 respectively. \textbf{(B) - (D)} Conductance sets before training \textbf{(B)}, after 2000 epochs \textbf{(C)} and after 10000 epochs \textbf{(D)}. Conductance spans from 0.0000869S to 0.0004416S, indicating final memristor RSs ranging from around 2264 ohms to 11507 ohms. \textbf{(E)} Concept diagram of spiking neural network used in this image recognition task. In practice the network is fully connected. \(22 \times 22\) -pixel images are unrolled to 484-bit input sending to 484 input neurons correspondingly as input spikes. In this task, a two-layer network with 484 input neurons and 10 output neurons is applied. \textbf{(F)} Accuracy curves over the training process for both memristor version and non-memristor version. Minibatch size of 100 are used to calculate the accuracy changes during the training. Provided 2000 images from a separated test set, memristor version and non-memristor version achieved accuracy of 82.00\% and 83.55\% respectively.}
    \label{fig:mnist}
\end{figure}

We use an MNIST handwritten digit recognition task as an example application to validate NeuroPack. Original images with \(28 \times 28\) pixels from the MNIST dataset have been cropped to \(22 \times 22\) and binarised (see input images for 10 digits in Figure \ref{fig:mnist} \textbf{(A)}). Background pixels and digit pixels are represented as 0 and 1 respectively. A single image is unrolled to a 484-dimensional vector with 0 and 1 only to be sent to input neurons in parallel. The spike encoding scheme sends a spike when the input bit is 1, and no spike for 0. The neural network used in this task is a 484-input 10-output two-layer feedforward winner-take-all spiking neural network with leaky-integrate-and-fire neuron model and gradient descent learning rule (see Figure \ref{fig:mnist} \textbf{(E)} for neural network architecture for this task). To map all 484\(\times\)10 synapses, a 100\(\times\)100 array of memristors is used. The network was trained for 10000 epochs before being fed with 2000 test images to evaluate classification accuracy. Parameters used in NeuroPack for MNIST handwritten digit classification task can be found in Table \ref{tab:mnist_params}. Memristor parameters listed in table are based on the extraction method from \citep{extraction} and devices presented in \citep{multibit}, given voltage range from 0.9V to 1.2V for positive bias and from -0.9V to -1.2V for negative bias with 11k$\Omega$ as initialised resistance. Therefore, pulse options used to update memristors are all within those ranges. The model yielded an estimated memristor operating range between $2.23\text{-}12.8\text{k}\Omega$ given bias voltage of $\pm1.2\text{V}$ and $12.5\text{-}18.9\text{k}\Omega$ given bias voltage of $\pm 0.9\text{V}$. The resulting conductance caused by the bias voltages of $\pm 0.9\text{V}\text{-}\pm1.2\text{V}$ is $5.3 \times 10^{-5} \text{-} 4.48 \times 10^{-4} \text{S} $. To make sure the weights can be mapped to range [0, 1], the linear mapping between memristor conductance and weights is given by the equation as below:
\[W = 2.53 \times10^3 \times G - 0.1337\]
Before training, memristor RSs are initialised to 11k$\Omega$ with a maximum variation of \(\pm\)500 $\Omega$. Memristor initial RSs are mapped to small weights close to the bottom boundary of the operating range, given conductance as the linear mapping of synaptic weights. Conductance sets before training, after 2000 epochs, and after 10000 epochs can be found in Figure \ref{fig:mnist} \textbf{(B)-(D)}. During training, conductance associated with digit pixels and labelled output neurons will be increased gradually, while other conductance stay small, therefore targeted digits show up in the conductance sets along the training process. The weight sets show the same tendency as the mapping is linear. Figure \ref{fig:mnist} \textbf{(F)} shows the accuracy evolution curve during training with minibatch size = 100. 2000 images from a separated test set were fed to the network after training, and 1640 out of 2000 got classified correctly, giving general inference accuracy of 82.00\%. The baseline given by the version storing weights directly without using memristor models achieved a test accuracy of 83.55\%, which indicates the accuracy bottleneck is not the memristor model.

\begin{figure}[t]
    \centering
    \includegraphics[width = 1.0\linewidth]{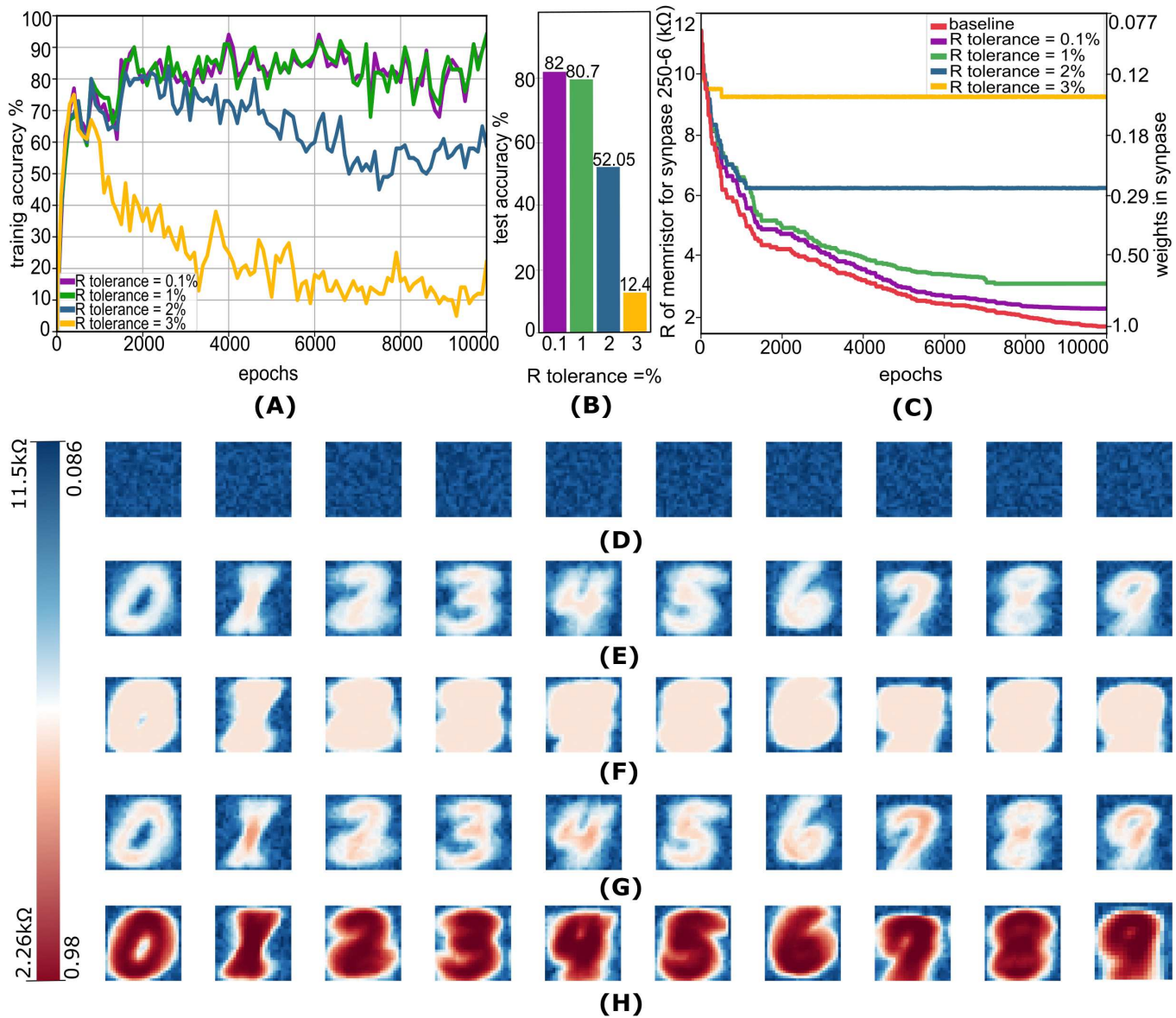}
    \caption{Classification accuracy affected by R tolerance. \textbf{(A)} and \textbf{(B)} Accuracy in training and test phases with different R tolerance respectively. \textbf{(C)} Resistance changes in a memristor representing a stimulated synapse (between input neuron 250 and output neuron 6) in training process with different R tolerance. Notice that we use virtual resistance values for the baseline. \textbf{(D)}-\textbf{(H)} Memristor resistive states before training \textbf{(D)}, after 1000 epochs for R tolerance of 2\%\textbf{(E)} or 0.1\%\textbf{(F)}, and after 10000 epochs for R tolerance of 2\%\textbf{(G)} or 0.1\%\textbf{(H)}. The total range of memristor RSs is from 2.26k\(\Omega\) to 11.5k\(\Omega\), which gives the range of weights as 0.98 to 0.086 correspondingly.}
    \label{fig:r_tolerance}
\end{figure}

We now use NeuroPack to illustrate how the intimately device- and programming protocol-related issue of selecting an appropriate R-tolerance affects recognition accuracy. Figure \ref{fig:r_tolerance} \textbf{(A)} and \textbf{(B)} shows the training accuracy curves and test set accuracy results for different R-tolerance values. When R-tolerance is small (within 1\%), the accuracy is not affected significantly. With a larger R-tolerance, training accuracy increases initially but then starts to drop. This is also reflected in the corresponding test set accuracy. To investigate the causes, we look into the resistance changes in both stimulated and non-stimulated synapses with different R-tolerance values, as displayed in Figure \ref{fig:r_tolerance} \textbf{(C)}. The red line shows the baseline virtual resistance values calculated according to the weight mapping scheme for a stimulated synapse (specifically the one between input neuron 250 and output neuron 6) as yielded by a non-memristor, software synapse. The baseline shows a gradual decrease tendency through out the whole 10000 training epochs. In contrast, resistance update with different R-tolerance values cuts off increasingly early as R-tolerance increases: for 0.1\%, 1\%, 2\%, and 3\% saturation occurs roughly at epochs \(\sim\) 9000, 7000, 1000, and 100 respectively. Intuitively speaking, if we use a smooth, continuous curve to fit the baseline trace, we find that its gradient progressively decreases. This can be explained by the decreasing gradient of the cost function during the training process. This indicates that the required resistance changes between successive time steps reduce as training continues. Meanwhile, R-tolerance is defined as \(\frac{R_{new} - R_{expected}}{R_{expected}}\) in NeuroPack, therefore it can be regarded as the cut-off ratio of memristor RS change. In other words, when resistance change between two time steps is smaller than the R-tolerance, the resistance update stops. Therefore, the larger the R-tolerance, the earlier memristor RSs stop updating. Figure \ref{fig:r_tolerance} \textbf{(D) - (H)} shows the memristor RS sets before training \textbf{(D)}, after 1000 epochs (with R tolerance of 2\% \textbf{(E)} and 0.1\% \textbf{(F)}), and after 10000 epochs (with R tolerance of 2\% \textbf{(G)} and 0.1\% \textbf{(H)}). There are three color regions that can be clearly seen: blue (high resistive range), white (middle resistive range), and red (low resistive range). Before training, memristor RSs are initialised to the high resistive range. After 1000 epochs, both versions with R tolerance of 2\% and 0.1\% with show distinguishable high and middle resistive ranges, reflecting the increasing training accuracy before 1000 epochs in Figure \ref{fig:r_tolerance} \textbf{(A)}. After 10000 epochs, the version with R-tolerance of 0.1\% clearly displays high, middle, and low resistive regions, while low resistive region is merged to middle region in version with R-tolerance of 2\% because the cut-off of a too large R-tolerance. The version with R-tolerance of 2\% is not able to distinguish specific images when middle region becomes larger as stimuli from different digits will all be assigned to weights with middle values. Therefore, the accuracy curves for large R-tolerance display decreasing tendency after certain points in Figure \ref{fig:r_tolerance} \textbf{(A)}. 

\begin{figure}[t]
    \centering
    \includegraphics[width = 1.0\linewidth]{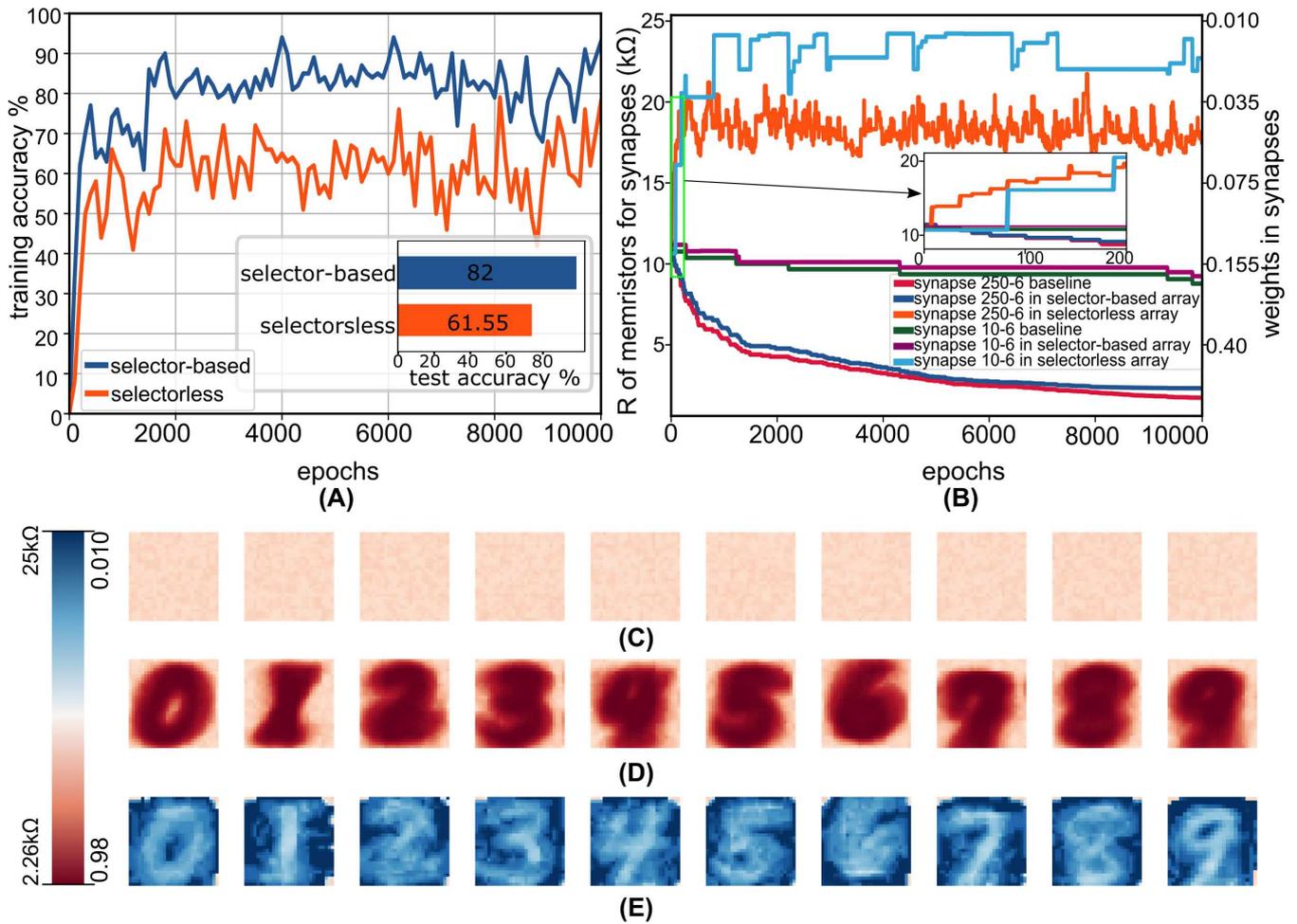}
    \caption{Classification with versions using selector-based and selectorless memristor arrays. \textbf{(A)} Accuracy in selector-based and selectorless arrays. \textbf{(B)} Resistance changes in memristors representing a stimulated (between input neuron 250 and output 6) and a non-stimulated (between input neuron 10 and output neuron 6) synapses. \textbf{(C) - (E)} give memristor resistive states before training \textbf{(C)}, after training for both selector-based \textbf{(D)} and selectorless \textbf{(E)} versions. The color bar shows the memristor states from 2.26k-25k$Omega$, which give weights in range 0.98 to 0.01 correspondingly.}
    \label{fig:selector}
\end{figure}

Finally, we present a comparison between two biasing schemes: a) bias voltages are only applied to selected devices (corresponding selector-based crossbar array) scenarios, and b) half-bias voltages are also applied to unselected devices in the same bitlines and wordlines (as in selectorless crossbars). Figure \ref{fig:selector} \textbf{(A)} shows the accuracy of both versions. In the training accuracy curves, both versions show the same tendency with a noticeable gap in between. The test accuracy bar chart further displays the\(~\sim\)20\% gap. In order to explore the cause of the accuracy gap, we look into the resistance change curves of memristors representing a stimulated synapse (between input neuron 250 and output neuron 6) and a non-stimulated synapse (between input neuron 10 and output neuron 6). The baseline curves (red) are given by virtual resistance values of the baseline which stores weights directly without using memristor models. The resistance change for the stimulated synapse in the selector-based version (dark blue trace) shows the same decreasing tendency as the baseline (red), while the resistance in the selectorless version (orange) displays an increasing tendency. Zooming into epochs 0 to 200, we observe unexpected resistance increases when no resistance update should happen for the selectorless scenario. This is because some memristors representing synapses connected to the neurons that are not supposed to fire have been applied pulses to increase the weights according to the learning rule, and they shared the same wordlines or bitlines with those whose resistance are supposed to decrease. A cycle of half-voltage bias only caused a trivial resistance increment, but there were many cycles of unexpected resistance drop happening in the same time step as there were many devices in the same wordlines/ bitlines, therefore the resistance increment accumulated and caused a large gap to the baseline. For the non-stimulated synapse, both the selector-based (purple) and the baseline (green) traces stay around their initial values through out the whole training phase, whilst unexpected resistance increases occur in the selectorless version. However, the resistance in memristors representing the stimulated synapse is still slightly smaller than that of the non-stimulated synapse in the given examples in the selectorless scenario. Because of this slight resistance difference, the NN based on selectorless array is still able to learn images and classify correctly in some cases. We present the memristor resistive states before and after training for both versions in Figure \ref{fig:selector} \textbf{(C) - (E)}. Due to the half-voltage bias, the new memeristor operation range changes to 2.23k-28k$\Omega$, giving the new conductance range from $3.57\times10^{-5}$S to $4.48\times10^{-4}$S. Therefore, the linear mapping from conductance to weights now is changed to the equation below:
\[W = 2.42 \times10^3 \times G - 0.0866\]
Memristor RSs are initialised as \(\sim\)11k\(\Omega\) before the training. After 10000 epochs, the resistive states of stimulated memristors in selector-based array (Figure \ref{fig:selector} \textbf{(D)}) decrease to low resistive range (\(\sim\)2.26k\(\Omega\)), while the non-stimulated stay in high resistive range (\(\sim\)11k\(\Omega\)). In the selectorless version, stimulated memristor RSs increase to very high resistive range (\(\sim\)18k\(\Omega\)) with non-stimulated ones increasing even higher to (\(\sim\)22k\(\Omega\)). Thus, NeuroPack has helped us uncover the perhaps surprising result that even in the presence of invasive unexpected weights update the NN is still capable of distinguishing the MNIST digits substantially better than chance, albeit with a very different weight distribution than the selector-based network.

\begin{table}[h]
    \centering
    \caption{Parameters used in NeuroPack for MNIST handwritten digit classification task.}
    \begin{tabular}{c c}
    
    \hline
    Global settings&\\
    \hline
    array size & 100\(\times\)100 \\
    array type & with selectors \\
    read noise & 0.1\% \\
    \hline
    Neuron model& \\
    \hline
    threshold &  25.16 or 24.16(for biasing method comparison only)\\
    leakage &  -0.3 \\
    \hline
    Learning rule& \\
    \hline
    learning rate & \(3.5 \times 10^{-6}\)\\
    noise scale & \(10^{-6}\)\\
    \hline
    Memristor model& \\
    \hline
    \(A_p\) & 0.21389\\
    \(A_n\) & -0.81302\\
    \(t_p\) & 1.6591\\
    \(t_n\) & 1.5148\\
    \(a_{0p}\) & 37087\\
    \(a_{0n}\) & 43430\\
    \(a_{1p}\) & -20193\\
    \(a_{1n}\) & 34333\\
    \hline     
    Weights updating & \\ 
    \hline
    voltage & \(\pm\)0.9, \(\pm\)1.1, \(\pm\)1.2, \(\pm\)1.2, \(\pm\)1.2, \(\pm\)1.2\\ 
    pulsewidth & \(10^{-6}\), \(10^{-6}\), \(10^{-6}\), \(5\times 10^{-6}\), \(10^{-5}\), \(5\times10^{-5}\)\\  
    R tolerance & 0.1\%\\
    max update steps & 5\\
    \hline
    \end{tabular}
    \label{tab:mnist_params}
\end{table}

\section{CONCLUSION}
In this paper, we presented NeuroPack, a versatile algorithm-level software emulator for memristor-based neuro-inspired computing systems. NeuroPack allows users to customise the simulator at both system level and device level. This platform can work as a standalone tool to emulate neuro-inspired computing with different neuron models, learning rules, memristor models, different types and numbers of memristor devices, different neural network architectures, and different applications. We further showcased an application example using NeuroPack to simulate a two-layer SNN for handwritten digit recognition with the MNIST dataset. We explored how different factors such as R-tolerance in weight updating and biasing methods in different array structures affect system classification accuracy and quickly reached two conclusions: a) That even a surprisingly lax 1\% tolerance in resistive state (an engineering parameter) allows for sufficient training efficiency to closely match the performance of an ideal model for this architecture and dataset. This indicates that memristor-based systems may be able to achieve competitive performance without requiring expensive precision circuits are least in some scenarios. b) That even in a scenario with unexpected weight updates due to the half-voltage biasing method in selectorless arrays it may be possible to decode useful information from the state of the system after training. These investigations illustrate the role NeuroPack can play in assisting users to design and validate neuro-inspired concepts and improve system performance involving emerging nanoscale memory technologies. We envisage that by varying datasets, biasing and other experimental parameters, device technology and connectivity patterns, users from across the community will be able to use this tool to generate the results that suit their needs quickly and efficiently.

\section*{Conflict of Interest Statement}
The authors declare that the research was conducted in the absence of any commercial or financial relationships that could be construed as a potential conflict of interest.

\section*{Funding}
The authors acknowledge the support of the EPSRC FORTE Programme Grant (EP/R024642/1), the RAEng Chair in Emerging Technologies (CiET1819/2/93), as well as the EU projects SYNCH (824162) and CHIST-ERA net SMALL.

\section*{Data Availability Statement}
The original code of this work can be found in https://github.com/hjq310/NeuroPack.

\bibliographystyle{frontiersinSCNS_ENG_HUMS} 
\bibliography{test}

\begin{thebibliography}{47}
\providecommand{\natexlab}[1]{#1}
\expandafter\ifx\csname urlstyle\endcsname\relax
  \providecommand{\doi}[1]{doi:\discretionary{}{}{}#1}\else
  \providecommand{\doi}{doi:\discretionary{}{}{}\begingroup
  \urlstyle{rm}\Url}\fi
\providecommand{\selectlanguage}[1]{\relax}
\providecommand{\bibAnnoteFile}[1]{%
  \IfFileExists{#1}{\begin{quotation}\noindent\textsc{Key:} #1\\
  \textsc{Annotation:}\ \input{#1}\end{quotation}}{}}
\providecommand{\bibAnnote}[2]{%
  \begin{quotation}\noindent\textsc{Key:} #1\\
  \textsc{Annotation:}\ #2\end{quotation}}

\bibitem[{Abbott(1999)}]{lif}
Abbott, L. (1999).
\newblock Lapicque’s introduction of the integrate-and-fire model neuron
  (1907).
\newblock \emph{Brain Research Bulletin} 50, 303--304.
\newblock \doi{https://doi.org/10.1016/S0361-9230(99)00161-6}
\bibAnnoteFile{lif}

\bibitem[{Akopyan et~al.(2015)Akopyan, Sawada, Cassidy, Alvarez-Icaza, Arthur,
  Merolla et~al.}]{truenorth}
Akopyan, F., Sawada, J., Cassidy, A., Alvarez-Icaza, R., Arthur, J., Merolla,
  P., et~al. (2015).
\newblock Truenorth: Design and tool flow of a 65 mw 1 million neuron
  programmable neurosynaptic chip.
\newblock \emph{IEEE Transactions on Computer-Aided Design of Integrated
  Circuits and Systems} 34, 1537--1557.
\newblock \doi{10.1109/TCAD.2015.2474396}
\bibAnnoteFile{truenorth}

\bibitem[{Aono and Hasegawa(2010)}]{AtomicSwitch}
Aono, M. and Hasegawa, T. (2010).
\newblock The atomic switch.
\newblock \emph{Proceedings of the IEEE} 98, 2228--2236.
\newblock \doi{10.1109/JPROC.2010.2061830}
\bibAnnoteFile{AtomicSwitch}

\bibitem[{Bedeschi et~al.(2009)Bedeschi, Fackenthal, Resta, Donze,
  Jagasivamani, Buda et~al.}]{pcm}
Bedeschi, F., Fackenthal, R., Resta, C., Donze, E.~M., Jagasivamani, M., Buda,
  E.~C., et~al. (2009).
\newblock A bipolar-selected phase change memory featuring multi-level cell
  storage.
\newblock \emph{IEEE Journal of Solid-State Circuits} 44, 217--227.
\newblock \doi{10.1109/JSSC.2008.2006439}
\bibAnnoteFile{pcm}

\bibitem[{Benjamin et~al.(2014)Benjamin, Gao, McQuinn, Choudhary,
  Chandrasekaran, Bussat et~al.}]{neurogrid}
Benjamin, B.~V., Gao, P., McQuinn, E., Choudhary, S., Chandrasekaran, A.~R.,
  Bussat, J.-M., et~al. (2014).
\newblock Neurogrid: A mixed-analog-digital multichip system for large-scale
  neural simulations.
\newblock \emph{Proceedings of the IEEE} 102, 699--716.
\newblock \doi{10.1109/JPROC.2014.2313565}
\bibAnnoteFile{neurogrid}

\bibitem[{Burr et~al.(2016)Burr, Brightsky, Sebastian, Cheng, Wu, Kim
  et~al.}]{pcm-ibm}
Burr, G.~W., Brightsky, M.~J., Sebastian, A., Cheng, H.-Y., Wu, J.-Y., Kim, S.,
  et~al. (2016).
\newblock Recent progress in phase-change memory technology.
\newblock \emph{IEEE Journal on Emerging and Selected Topics in Circuits and
  Systems} 6, 146--162.
\newblock \doi{10.1109/JETCAS.2016.2547718}
\bibAnnoteFile{pcm-ibm}

\bibitem[{Chen et~al.(2018)Chen, Peng, and Yu}]{neurosim}
Chen, P.-Y., Peng, X., and Yu, S. (2018).
\newblock Neurosim: A circuit-level macro model for benchmarking neuro-inspired
  architectures in online learning.
\newblock \emph{IEEE Transactions on Computer-Aided Design of Integrated
  Circuits and Systems} 37, 3067--3080.
\newblock \doi{10.1109/TCAD.2018.2789723}
\bibAnnoteFile{neurosim}

\bibitem[{Choi et~al.(2016)Choi, Torrezan, Strachan, Kotula, Lohn, Marinella
  et~al.}]{highspeedrram}
Choi, B.~J., Torrezan, A., Strachan, J.~W., Kotula, P., Lohn, A., Marinella,
  M., et~al. (2016).
\newblock High-speed and low-energy nitride memristors.
\newblock \emph{Advanced Functional Materials} 26.
\newblock \doi{10.1002/adfm.201600680}
\bibAnnoteFile{highspeedrram}

\bibitem[{Chua(1971)}]{memristor}
Chua, L. (1971).
\newblock Memristor-the missing circuit element.
\newblock \emph{IEEE Transactions on Circuit Theory} 18, 507--519.
\newblock \doi{10.1109/TCT.1971.1083337}
\bibAnnoteFile{memristor}

\bibitem[{Covi et~al.(2016)Covi, Brivio, Serb, Prodromakis, Fanciulli, and
  Spiga}]{unsupervised2}
Covi, E., Brivio, S., Serb, A., Prodromakis, T., Fanciulli, M., and Spiga, S.
  (2016).
\newblock Analog memristive synapse in spiking networks implementing
  unsupervised learning.
\newblock \emph{Frontiers in Neuroscience} 10, 482.
\newblock \doi{10.3389/fnins.2016.00482}
\bibAnnoteFile{unsupervised2}

\bibitem[{Davies et~al.(2018)Davies, Srinivasa, Lin, Chinya, Cao, Choday
  et~al.}]{loihi}
Davies, M., Srinivasa, N., Lin, T.-H., Chinya, G., Cao, Y., Choday, S.~H.,
  et~al. (2018).
\newblock Loihi: A neuromorphic manycore processor with on-chip learning.
\newblock \emph{IEEE Micro} 38, 82--99.
\newblock \doi{10.1109/MM.2018.112130359}
\bibAnnoteFile{loihi}

\bibitem[{Dundar and Rose(1995)}]{signalQuant}
Dundar, G. and Rose, K. (1995).
\newblock The effects of quantization on multilayer neural networks.
\newblock \emph{IEEE Transactions on Neural Networks} 6, 1446--1451.
\newblock \doi{10.1109/72.471364}
\bibAnnoteFile{signalQuant}

\bibitem[{Frenkel et~al.(2021)Frenkel, Lefebvre, and Bol}]{drtp}
Frenkel, C., Lefebvre, M., and Bol, D. (2021).
\newblock Learning without feedback: Fixed random learning signals allow for
  feedforward training of deep neural networks.
\newblock \emph{Frontiers in Neuroscience} 15, 20.
\newblock \doi{10.3389/fnins.2021.629892}
\bibAnnoteFile{drtp}

\bibitem[{Gerstner et~al.(2014)Gerstner, Kistler, Naud, and
  Paninski}]{originalLIF}
Gerstner, W., Kistler, W.~M., Naud, R., and Paninski, L. (2014).
\newblock Neuronal dynamics: From single neurons to networks and models of
  cognition
\bibAnnoteFile{originalLIF}

\bibitem[{Goux et~al.(2012)Goux, Fantini, Kar, Chen, Jossart, Degraeve
  et~al.}]{ultralowcurrentrram}
Goux, L., Fantini, A., Kar, G., Chen, Y.-Y., Jossart, N., Degraeve, R., et~al.
  (2012).
\newblock Ultralow sub-500na operating current high-performance tin
  $\backslash$ al2o3 $\backslash$ hfo2 $\backslash$ hf $\backslash$ tin bipolar
  rram achieved through understanding-based stack-engineering.
\newblock In \emph{2012 Symposium on VLSI Technology (VLSIT)}. 159--160.
\newblock \doi{10.1109/VLSIT.2012.6242510}
\bibAnnoteFile{ultralowcurrentrram}

\bibitem[{Guo et~al.(2019)Guo, Wu, Gao, and Qian}]{memristor_snn_unsupervised}
Guo, Y., Wu, H., Gao, B., and Qian, H. (2019).
\newblock {Unsupervised learning on resistive memory array based spiking neural
  networks}.
\newblock \emph{Frontiers in Neuroscience} 13, 1--16.
\newblock \doi{10.3389/fnins.2019.00812}
\bibAnnoteFile{memristor_snn_unsupervised}

\bibitem[{G{\"u}tig and Sompolinsky(2006)}]{tempotron}
G{\"u}tig, R. and Sompolinsky, H. (2006).
\newblock The tempotron: a neuron that learns spike timing–based decisions.
\newblock \emph{Nature Neuroscience} 9, 420--428
\bibAnnoteFile{tempotron}

\bibitem[{Hochreiter and Schmidhuber(1997)}]{lstm}
Hochreiter, S. and Schmidhuber, J. (1997).
\newblock Long short-term memory.
\newblock \emph{Neural Comput.} 9, 1735–1780.
\newblock \doi{10.1162/neco.1997.9.8.1735}
\bibAnnoteFile{lstm}

\bibitem[{Hu et~al.(2018)Hu, Graves, Li, Li, Ge, Montgomery
  et~al.}]{dotproduct}
Hu, M., Graves, C.~E., Li, C., Li, Y., Ge, N., Montgomery, E., et~al. (2018).
\newblock Memristor-based analog computation and neural network classification
  with a dot product engine.
\newblock \emph{Advanced Materials}
\bibAnnoteFile{dotproduct}

\bibitem[{Izhikevich(2003)}]{Izhikevich}
Izhikevich, E. (2003).
\newblock Simple model of spiking neurons.
\newblock \emph{IEEE Transactions on Neural Networks} 14, 1569--1572.
\newblock \doi{10.1109/TNN.2003.820440}
\bibAnnoteFile{Izhikevich}

\bibitem[{Khiat et~al.(2016)Khiat, Ayliffe, and Prodromakis}]{extremeScale}
Khiat, A., Ayliffe, P., and Prodromakis, T. (2016).
\newblock High density crossbar arrays with sub- 15 nm single cells via liftoff
  process only.
\newblock \emph{Scientific Reports} 6, 1--8
\bibAnnoteFile{extremeScale}

\bibitem[{LeCun et~al.(1999)LeCun, Haffner, Bottou, and Bengio}]{cnn}
LeCun, Y., Haffner, P., Bottou, L., and Bengio, Y. (1999).
\newblock Object recognition with gradient-based learning.
\newblock In \emph{Shape, Contour and Grouping in Computer Vision}
\bibAnnoteFile{cnn}

\bibitem[{Lee et~al.(2016)Lee, Delbruck, and
  Pfeiffer}]{backpropagation_with_noise}
Lee, J.~H., Delbruck, T., and Pfeiffer, M. (2016).
\newblock Training deep spiking neural networks using backpropagation.
\newblock \emph{Frontiers in Neuroscience} 10, 508.
\newblock \doi{10.3389/fnins.2016.00508}
\bibAnnoteFile{backpropagation_with_noise}

\bibitem[{Li et~al.(2018)Li, Belkin, Li, Yan, Hu, Ge et~al.}]{insitu}
Li, C., Belkin, D., Li, Y., Yan, P., Hu, M., Ge, N., et~al. (2018).
\newblock Efficient and self-adaptive in-situ learning in multilayer memristor
  neural networks
\bibAnnoteFile{insitu}

\bibitem[{Markram et~al.(1997)Markram, Lübke, Frotscher, and Sakmann}]{stdp}
Markram, H., Lübke, J., Frotscher, M., and Sakmann, B. (1997).
\newblock Regulation of synaptic efficacy by coincidence of postsynaptic aps
  and epsps.
\newblock \emph{Science (New York, N.Y.)} 275, 213--5.
\newblock \doi{10.1126/science.275.5297.213}
\bibAnnoteFile{stdp}

\bibitem[{Merolla et~al.(2011)Merolla, Arthur, Akopyan, Imam, Manohar, and
  Modha}]{digital_crossbar}
Merolla, P., Arthur, J., Akopyan, F., Imam, N., Manohar, R., and Modha, D.
  (2011).
\newblock A digital neurosynaptic core using embedded crossbar memory with 45pj
  per spike in 45nm.
\newblock 1 -- 4.
\newblock \doi{10.1109/CICC.2011.6055294}
\bibAnnoteFile{digital_crossbar}

\bibitem[{Messaris et~al.(2017{\natexlab{a}})Messaris, Nikolaidis, Serb,
  Stathopoulos, Gupta, Khiat et~al.}]{extraction}
Messaris, Y., Nikolaidis, S., Serb, A., Stathopoulos, S., Gupta, I., Khiat, A.,
  et~al. (2017{\natexlab{a}}).
\newblock A tio2 reram parameter extraction method.
\newblock 1--4.
\newblock \doi{10.1109/ISCAS.2017.8050789}
\bibAnnoteFile{extraction}

\bibitem[{Messaris et~al.(2017{\natexlab{b}})Messaris, Serb, Khiat, Nikolaidis,
  and Prodromakis}]{rrammodel}
Messaris, Y., Serb, A., Khiat, A., Nikolaidis, S., and Prodromakis, T.
  (2017{\natexlab{b}}).
\newblock A compact verilog-a reram switching model
\bibAnnoteFile{rrammodel}

\bibitem[{Oster et~al.(2009)Oster, Douglas, and Liu}]{wta}
Oster, M., Douglas, R.~J., and Liu, S.-C. (2009).
\newblock Computation with spikes in a winner-take-all network.
\newblock \emph{Neural Computation} 21, 2437--2465
\bibAnnoteFile{wta}

\bibitem[{Painkras et~al.(2013)Painkras, Plana, Garside, Temple, Galluppi,
  Patterson et~al.}]{spinnaker}
Painkras, E., Plana, L.~A., Garside, J., Temple, S., Galluppi, F., Patterson,
  C., et~al. (2013).
\newblock Spinnaker: A 1-w 18-core system-on-chip for massively-parallel neural
  network simulation.
\newblock \emph{IEEE Journal of Solid-State Circuits} 48, 1943--1953.
\newblock \doi{10.1109/JSSC.2013.2259038}
\bibAnnoteFile{spinnaker}

\bibitem[{Payvand et~al.(2020)Payvand, Fouda, Kurdahi, Eltawil, Member, and
  Neftci}]{online}
Payvand, M., Fouda, M.~E., Kurdahi, F., Eltawil, A.~M., Member, S., and Neftci,
  E.~O. (2020).
\newblock {On-Chip Error-Triggered Learning of Multi-Layer Memristive Spiking
  Neural Networks} 10, 522--535
\bibAnnoteFile{online}

\bibitem[{Prodromakis et~al.(2010)Prodromakis, Michelakis, and
  Toumazou}]{metal-oxide}
Prodromakis, T., Michelakis, K., and Toumazou, C. (2010).
\newblock Switching mechanisms in microscale memristors.
\newblock \emph{Electronics Letters} 46, 63 -- 65.
\newblock \doi{10.1049/el.2010.2716}
\bibAnnoteFile{metal-oxide}

\bibitem[{Rumelhart et~al.(1986)Rumelhart, Hinton, and Williams}]{bp}
Rumelhart, D., Hinton, G.~E., and Williams, R.~J. (1986).
\newblock Learning representations by back-propagating errors.
\newblock \emph{Nature} 323, 533--536
\bibAnnoteFile{bp}

\bibitem[{Schmitt et~al.(2017)Schmitt, Klahn, Bellec, Grubl, Guttler, Hartel
  et~al.}]{brainscales}
Schmitt, S., Klahn, J., Bellec, G., Grubl, A., Guttler, M., Hartel, A., et~al.
  (2017).
\newblock Neuromorphic hardware in the loop: Training a deep spiking network on
  the brainscales wafer-scale system.
\newblock \emph{2017 International Joint Conference on Neural Networks (IJCNN)}
  \doi{10.1109/ijcnn.2017.7966125}
\bibAnnoteFile{brainscales}

\bibitem[{Sengupta et~al.(2019)Sengupta, Ye, Wang, Liu, and Roy}]{SNN}
Sengupta, A., Ye, Y., Wang, R., Liu, C., and Roy, K. (2019).
\newblock Going deeper in spiking neural networks: Vgg and residual
  architectures.
\newblock \emph{Frontiers in Neuroscience} 13, 95.
\newblock \doi{10.3389/fnins.2019.00095}
\bibAnnoteFile{SNN}

\bibitem[{Serb et~al.(2016)Serb, Bill, Khiat, Berdan, Legenstein, and
  Prodromakis}]{unsupervised1}
Serb, A., Bill, J., Khiat, A., Berdan, R., Legenstein, R., and Prodromakis, T.
  (2016).
\newblock Unsupervised learning in probabilistic neural networks with
  multi-state metal-oxide memristive synapses.
\newblock \emph{Nature Communications} 7.
\newblock \doi{doi:10.1038/ncomms12611}
\bibAnnoteFile{unsupervised1}

\bibitem[{Serb et~al.(2017)Serb, Manino, Messaris, Tran-Thanh, and
  Prodromakis}]{hardwarebayesian}
Serb, A., Manino, E., Messaris, I., Tran-Thanh, L., and Prodromakis, T. (2017).
\newblock Hardware-level bayesian inference.
\newblock In \emph{Neural Information Processing Systems}
\bibAnnoteFile{hardwarebayesian}

\bibitem[{Serrano-Gotarredona et~al.(2013)Serrano-Gotarredona, Masquelier,
  Prodromakis, Indiveri, and Linares-Barranco}]{stdpandrram}
Serrano-Gotarredona, T., Masquelier, T., Prodromakis, T., Indiveri, G., and
  Linares-Barranco, B. (2013).
\newblock Stdp and stdp variations with memristors forspiking neuromorphic
  learning systems.
\newblock \emph{Frontiers in Neuroscience} 7, 1--15
\bibAnnoteFile{stdpandrram}

\bibitem[{Shin et~al.(2010)Shin, Kim, and Kang}]{charge-flux}
Shin, S., Kim, K., and Kang, S.-M. (2010).
\newblock Compact models for memristors based on charge-flux constitutive
  relationships.
\newblock \emph{Computer-Aided Design of Integrated Circuits and Systems, IEEE
  Transactions on} 29, 590 -- 598.
\newblock \doi{10.1109/TCAD.2010.2042891}
\bibAnnoteFile{charge-flux}

\bibitem[{Sivan et~al.(2019)Sivan, Li, Veluri, Zhao, Tang, Wang et~al.}]{1t1r}
Sivan, M., Li, Y., Veluri, H., Zhao, Y., Tang, B., Wang, X., et~al. (2019).
\newblock All wse2 1t1r resistive ram cell for future monolithic 3d embedded
  memory integration.
\newblock \emph{Nature Communications} 10.
\newblock \doi{10.1038/s41467-019-13176-4}
\bibAnnoteFile{1t1r}

\bibitem[{Stathopoulos et~al.(2017)Stathopoulos, Khiat, Trapatseli, Cortese,
  Serb, Valov et~al.}]{multibit}
Stathopoulos, S., Khiat, A., Trapatseli, M., Cortese, S., Serb, A., Valov, I.,
  et~al. (2017).
\newblock Multibit memory operation of metal-oxide bi-layer memristors.
\newblock \emph{Scientific Reports} 7.
\newblock \doi{10.1038/s41598-017-17785-1}
\bibAnnoteFile{multibit}

\bibitem[{Vincent et~al.(2015)Vincent, Larroque, Locatelli, Ben~Romdhane,
  Bichler, Gamrat et~al.}]{stt}
Vincent, A.~F., Larroque, J., Locatelli, N., Ben~Romdhane, N., Bichler, O.,
  Gamrat, C., et~al. (2015).
\newblock Spin-transfer torque magnetic memory as a stochastic memristive
  synapse for neuromorphic systems.
\newblock \emph{IEEE Transactions on Biomedical Circuits and Systems} 9,
  166--174.
\newblock \doi{10.1109/TBCAS.2015.2414423}
\bibAnnoteFile{stt}

\bibitem[{Wu and Feng(2018)}]{ANN}
Wu, Y.-c. and Feng, J.-w. (2018).
\newblock Development and application of artificial neural network.
\newblock \emph{Wireless Personal Communications} 102, 1645--1656.
\newblock \doi{10.1007/s11277-017-5224-x}
\bibAnnoteFile{ANN}

\bibitem[{Xia et~al.(2018)Xia, Li, Tang, Gu, Chen, Yu et~al.}]{mnsim}
Xia, L., Li, B., Tang, T., Gu, P., Chen, P.-Y., Yu, S., et~al. (2018).
\newblock Mnsim: Simulation platform for memristor-based neuromorphic computing
  system.
\newblock \emph{IEEE Transactions on Computer-Aided Design of Integrated
  Circuits and Systems} 37, 1009--1022.
\newblock \doi{10.1109/TCAD.2017.2729466}
\bibAnnoteFile{mnsim}

\bibitem[{Xia et~al.(2009)Xia, Robinett, Cumbie, Banerjee, Cardinali, Yang
  et~al.}]{memristor-cmos}
Xia, Q., Robinett, W., Cumbie, M.~W., Banerjee, N., Cardinali, T.~J., Yang,
  J.~J., et~al. (2009).
\newblock Memristor-cmos hybrid integrated circuits for reconfigurable logic.
\newblock \emph{Nano Letters} 9, 3640--3645.
\newblock \doi{10.1021/nl901874j}
\bibAnnoteFile{memristor-cmos}

\bibitem[{Yao et~al.(2020)Yao, Wu, Gao, Tang, Zhang, Zhang et~al.}]{offline}
Yao, P., Wu, H., Gao, B., Tang, J., Zhang, Q., Zhang, W., et~al. (2020).
\newblock {Fully hardware-implemented memristor convolutional neural network}.
\newblock \emph{Nature} 577, 641--646.
\newblock \doi{10.1038/s41586-020-1942-4}
\bibAnnoteFile{offline}

\bibitem[{Yin et~al.(2017)Yin, Venkataramanaiah, Chen, Krishnamurthy, Cao,
  Chakrabarti et~al.}]{algorithm_and_hardware}
Yin, S., Venkataramanaiah, S.~K., Chen, G.~K., Krishnamurthy, R., Cao, Y.,
  Chakrabarti, C., et~al. (2017).
\newblock Algorithm and hardware design of discrete-time spiking neural
  networks based on back propagation with binary activations.
\newblock \emph{CoRR} abs/1709.06206
\bibAnnoteFile{algorithm_and_hardware}

\end{thebibliography}


\end{document}


\onecolumn
\firstpage{1}

\title[Supplementary Material]{{\helveticaitalic{Supplementary Material}}}

\maketitle

\section{derivation of back propagation for Leaky integrate-and-fire neuron models (without winner-take-all)}

Leaky integrate-and-fire neuron model in matrix form:

\[V_t = Wx_t + \alpha V_{t-1}\odot (1 - y_{t-1})\]
\[y_t = h(V_t - V_{th})\]
\[E = \frac{1}{2N}\sum_{i = 0}^{N}(y_{i,t} - \hat{y_{i,t}})^{2}\] 
\[\Delta W = - \eta \frac{\partial E}{\partial W}\]

Where $V_t$ is membrane voltage vector in the selected layer at time $t$ with the size of $n \times 1$ (n is the total number of neurons in this layer), $x_t$ is the input signals fed to selected layer at time $t$ with the size of $m \times 1$ (m is the total number of neurons in last layer), $W$ represents the weight matrix between current current layer and the last layer with the size of $n \times m$, $\alpha$ is the leakage factor, $\odot$ means the element-wise product, $V_{th}$ is the membrane voltage threshold, $y_t$ is the output vector for the selected layer with only 0 and 1 to indicate if there is a spike generated with the size of $n \times 1 $, and $\hat{y_t}$ is the correct output state for output layer with the same size of that of the output layer, $E$ is the cost function calculating the distance between $y_t$ in the output layer and $\hat{y_t}$, $N$ is the number of output neurons, and $\Delta W$ gives the weight update matrix with the same size as that of $W$.

According to chain rule, $\frac{\partial E}{\partial w_{l, t}}$ for the output layer can be derived as below:

\begin{align*}
    \frac{\partial E}{\partial w_{l, t}} &= \frac{\partial E}{\partial y_{l, t}}\cdot  \frac{\partial y_{l, t}}{\partial V_{l, t}}\cdot  \frac{\partial V_{l, t}}{\partial W_{l, t}} \\
    &= (y_{l,t} - \hat{y_{t}}) \odot h'(V_{l, t} - V_{th})\cdot x^T_{l, t}\\
    &= \delta_{l,t}x^T_{l, t} \quad \textbf{let } \delta_{l,t} = (y_{l,t} - \hat{y_{t}}) \odot h'(V_{l, t} - V_{th})
\end{align*}

After that we can back propagate the error from output layer:

\begin{align*}
    \frac{\partial E}{\partial w_{l-1, t}} &= \frac{\partial E}{\partial y_{l, t}}\cdot  \frac{\partial y_{l, t}}{\partial V_{l, t}}\cdot  \frac{\partial V_{l, t}}{\partial W_{l-1, t}} \\
    &= (y_{l,t} - \hat{y_{t}}) \odot h'(V_{l, t} - V_{th})\cdot \frac{\partial V_{l, t}}{\partial W_{l-1, t}}\\
    &= (W^T_{l, t}\delta_{l,t} \odot h'(V_{l-1, t} - V_{th}))x^T_{l-1, t} \\
    &= \delta_{l-1,t}x^T_{l-1, t} \quad \textbf{let } \delta_{l-1,t} = (W^T_{l, t}\delta_{l,t} \odot h'(V_{l-1, t} - V_{th}))
\end{align*}

Therefore, the weight update matrix $\Delta W$ can be expressed as follow:

\begin{align*}
    \Delta W_k &= -\eta\delta_{k, t}x_{k, t}^T \\
    \delta_{k, t} &= 
    \begin{cases}
    \frac{1}{N}(y_{k, t} - \hat{y}_t) \odot h'(V_{k, t} - V_{th}) & \textbf{if k = K}\\
    (W_{k+1, t}^T\delta_{k+1, t}) \odot h'(V_{k, t} - V_{th}) & \textbf{otherwise}
    \end{cases}
\end{align*}

Where $K$ is the index of the output layer.

\newpage
\section{derivation of back propagation for Leaky integrate-and-fire neuron models (with winner-take-all)}

To have winner-take-all network-level restriction to only allow at most one neuron to fire at each time step, a softmax layer is added to the previous output layer:

\[S_t = softmax(V_t \odot y_t)\]

Where $S_t$ is a vector with the size of $N \times 1$. With the new output representation becoming continuous values, we apply cross entropy loss as the cost function:

\[E = - \sum_{i=0}^{N}\hat{y_{i,t}}ln(S_{i,t}) = - ln(S_{j, t})\]

Where j is the index of the output neuron that should fire according to the ground truth.

Therefore, the derivation of the gradient for the output layer is as below:

\begin{align*}
    \frac{\partial E}{\partial w_{l, t}} &= \frac{\partial E}{\partial S_{t}}\cdot  \frac{\partial S_{t}}{\partial V_{l, t}}\cdot  \frac{\partial V_{l, t}}{\partial W_{l, t}} \\
    &= \frac{\partial E}{\partial S_{t}}\cdot  \frac{\partial S_{t}}{\partial Z_{l, t}}\cdot
    \frac{\partial Z_{l, t}}{\partial V_{l, t}}\cdot
    \frac{\partial V_{l, t}}{\partial W_{l, t}} \quad \textbf{let } Z_{l, t} = V_{l, t} \odot y_{l, t}\\
    &= (S_{t} - \hat{y_t}) \odot (y_{l, t} + V_{l, t} \odot h'(V_{l, t} - V_{th}) x^T_{l, t}\\
    &= \delta_{l, t}x^T_{l, t} \quad \textbf{let }\delta_{l, t} = (S_{t} - \hat{y_t}) \odot (y_{l, t} + V_{l, t} \odot h'(V_{l, t} - V_{th})\\ 
\end{align*}

From the output layer results, we can back propagate the errors and get the expression for other layers as follow:

\begin{align*}
    \frac{\partial E}{\partial w_{l-1, t}} &= \frac{\partial E}{\partial S_{t}}\cdot  \frac{\partial S_{t}}{\partial V_{l, t}}\cdot  \frac{\partial V_{l, t}}{\partial W_{l-1, t}} \\
    &= \frac{\partial E}{\partial S_{t}}\cdot  \frac{\partial S_{t}}{\partial Z_{l, t}}\cdot
    \frac{\partial Z_{l, t}}{\partial V_{l, t}}\cdot
    \frac{\partial V_{l, t}}{\partial W_{l-1, t}} \quad \textbf{let } Z_{l, t} = V_{l, t} \odot y_{l, t}\\
    &= (W^T_{l, t}\delta_{l, t} \odot h'(V_{l-1, t} - V_{th})) x^T_{l-1, t}\\
    &= \delta_{l-1, t}x^T_{l-1, t} \quad \textbf{let }\delta_{l, t} = ((W^T_{l, t}\delta_{l, t} \odot h'(V_{l-1, t} - V_{th}))\\ 
\end{align*}

Therefore, the final weight update matrix expression is as below:

\begin{align*}
    \Delta W_k &= -\eta\delta_{k, t}x_{k, t}^T \\
    \delta_{k, t} &= 
    \begin{cases}
    (S_{t} - \hat{y_t}) \odot (y_t + V_t \odot h'(V_{k, t} - V_{th})) & \textbf{if k = K}\\
    (W_{k+1, t}^T\delta_{k+1, t}) \odot h'(V_{k, t} - V_{th}) & \textbf{otherwise}
    \end{cases}
\end{align*}

\newpage
\section{Pseudo code for pulse parameter selection module}
\begin{algorithm}[h!]
    \begin{algorithmic} 
    \Procedure{Pulse parameters selection according to \(\Delta\)W}{}
        \State \(\text{memristor}=\text{\textbf{ParametericDevice}(*args)}\) \Comment{Create a \textbf{ParametericDevice} object}
        \State  \(\text{res}\) = [] \Comment{Create an empty list to store resulting resistance}
        \For{pulse in pulseList} \Comment{Loop over all options of pulse parameters}
            \State  \(\text{memristor.\textbf{initialise}}\)(R) \Comment{Set current resistance to \textbf{R}}
            \For{timestep in range(pulse[1] / dt)} \Comment{\textbf{pulse[1]} stores pulsewidth}
               \State  \textbf{step\_dt}(pulse[0], dt) \Comment{\textbf{pulse[0]} stores magnitude}   
            \EndFor
        \State  res.append(memristor.Rmem) \Comment{Append resulting resistance to the list}
        \EndFor
        \State  resDist = abs(res - R\_expected) \Comment{Calculate distance between expected and calculated resistance}
        \State  selectedPulseIndex = argmin(resDist) \Comment{Find the index of minimal distance}
        \State  selectedPulse = res[selectedPulseIndex] \Comment{Find parameters of the pulse option}
    \EndProcedure
    \end{algorithmic} 
\end{algorithm}
\newpage
